\documentclass[11pt]{article}

\usepackage{graphicx,amsfonts,amsbsy, amssymb, amsmath, mathrsfs}

\usepackage{amsthm}
\newtheorem{theorem}{Theorem}[section]
\newtheorem{proposition}[theorem]{Proposition}
\newtheorem{corollary}[theorem]{Corollary}
\newtheorem{lemma}[theorem]{Lemma}

\theoremstyle{definition}

\newtheorem{remark}[theorem]{Remark}

\renewcommand{\vec}[1]{{\bf{#1}}}

\newcommand{\curlyL}{{\mathscr L}}
\newcommand{\x}{\vec{x}}
\renewcommand{\L}{\vec{L}} 
\newcommand{\diag}{\mathop{\rm diag}}
\newcommand\cB{{\cal B}}
\newcommand\cV{{\cal V}}
\renewcommand{\S}{S}
\newcommand{\Id}{I}
\newcommand{\U}{U}
\newcommand{\A}{A}
\newcommand{\D}{D}
\newcommand{\bfT}{T}
\newcommand{\Lbar}{\bar{L}}

\newcommand{\I}{{\mathbb I}}
\newcommand{\R}{{\mathbb R}}
\newcommand{\Q}{{\mathbb Q}}
\newcommand{\Z}{{\mathbb Z}}

\newcommand{\Torus}{{{\mathbb T}^\numb}}

\newcommand{\rme}{{\mathrm e}}
\newcommand{\rmi}{{\mathrm i}}
\newcommand{\rmd}{{\mathrm d}}

\newcommand{\coloneq}{\mathbin{\hbox{\raise0.08ex\hbox{\rm :}}\!\!=}}

\newcommand{\that}{\hat{\theta}}

\DeclareMathOperator{\Ord}{{\rm O}}

\newcommand{\spec}{\lambda}
\newcommand{\Spec}{\Lambda}

\newcommand{\flow}{\phi}

\newcommand{\numb}{B}

\newcommand{\efun}{u}
\newcommand{\evS}{\boldsymbol{\phi}}
\newcommand{\evU}{\boldsymbol{\psi}}

\renewcommand{\dim}{{2\numb}}

\newcommand{\Daverage}[1]{{\mathbb E}^D\!\left( #1 \right)}

\newcommand{\vecalpha}{\boldsymbol{\alpha}}
\newcommand{\vecxi}{\boldsymbol{\xi}}

\begin{document}
\title{Relationship between scattering matrix and spectrum of quantum graphs}
\author{G. Berkolaiko${}^\clubsuit$ and B.\ Winn$^{\clubsuit,\spadesuit}$\\
{\protect\small ${}^\clubsuit$ Department of Mathematics, 
Texas A\&{M} University, College Station, {\sc Texas} 77843, USA. }\\
{\small ${}^\spadesuit$ School of Mathematics, Loughborough University,
Loughborough, LE11 3TU, UK}}
\date{19${}^{\rm th}$ February 2008}
\maketitle
\begin{abstract}
We investigate the equivalence between spectral characteristics of
the Laplace operator on a metric graph, and the associated unitary
scattering operator. We prove that the statistics of level spacings,
and moments of observations in the eigenbases coincide in the limit
that all bond lengths approach a positive constant value.
\end{abstract}

\thispagestyle{empty}

\section{Introduction}
\label{sec:intro}

Quantum graphs have attracted much attention in recent years due both
to their applicability as physical models, and their interesting 
mathematical properties. We refer the reader to \cite{ber:qgt} and
the forthcoming volume \cite{exn:aga} for a
{\it pot-pourri}\/ of new results.  Recent reviews, dedicated to quantum
graphs, include \cite{kuc:gmw, GS06}.

In this article we focus on one feature of interest which is the use
of graph models to probe the 
the universality of quantum systems. One of the unsolved paradoxes
of quantum mechanics is the observation that a great many quantum
systems are remarkably similar when one makes statistical
observations in the semi-classical r\'egime. This manifests itself
both in the energy levels, and associated energy eigenfunctions.
Despite a great deal of effort, this universality is poorly understood
mathematically.
Generic quantum graphs exhibit this universal behavior, and represent
the most likely system for which a full mathematically rigorous proof
of this universality will first be found.  Important steps in this
direction have been taken in \cite{GA05}.

A {\it quantum graph} can be defined in two different, but
related, ways (a complete description appears in the following section).  
One may consider a self-adjoint realisation of the Laplace operator,
or a scattering matrix approach. Mathematically, the scattering approach
appears to be more tractable, and has formed the basis of most
investigations \cite{sch:ssq,Tan01,BSW02,ber:ffe,ber:qeg}. 
Moreover, the spectra arising from these 
quantizations are subtly different.
This is not as confusing as it might seem, since the statistical
properties of both versions of the spectrum are believed to coincide
when averaged over a large interval.
It is the purpose of this article to put a concrete mathematical 
foundation behind this belief.

The plan of the article is as follows: In the next section we
give precise definitions of the two ways to describe quantum
graphs, and in section \ref{sec:results} describe our main
results. In section \ref{sec:tools} we describe the tools
used and then present in section \ref{sec:cinque} the proofs
of our results.

\section{Two descriptions of quantum graphs} \label{sec:due}

For both constructions of a quantum graph we begin with a graph $G =
(\cV,\cB)$ where $\cV$ is a finite set of vertices (sometimes referred to
as nodes), and
$\cB$ is the set of bonds (or edges).  Each bond $b$ has a
positive length, denoted $L_b$.  The total number of bonds is $\numb$.
We denote by $d_v$ the degree of the vertex $v\in\cV$, which is
the number of bonds emanating from it.

\subsection{The Laplace operator approach}
The first way to define a quantum graph is to identify each bond
$b$ with the interval $[0,L_b]$ of the real line and thus define the
$L^2$-space of functions on the graph.  Then one can consider the eigenproblem
\begin{equation}
  \label{eq:shrod_eq}
  -\frac{\rmd^2}{\rmd x^2} \efun_b(x) = \spec^2\efun_b(x).
\end{equation}
This setup has a long history of being used in physical models
\cite{Pau36,Gri53,RS53}.  It was studied by mathematicians since at least the
1980s \cite{Lum80,Roth83b,vB85,Nic87,PP88,ger:asp,exn:fqm}.  
The study of spectral
statistics of quantum graphs was initiated in \cite{KS97,KS99}.

To make the operator in (\ref{eq:shrod_eq}) self-adjoint one needs to impose
matching conditions on the behavior of $\efun$ at the vertices of the graph.
One possibility is to impose Kirchhoff conditions:\footnote{sometimes called
  ``Neumann'' conditions} we require that $\efun$ is continuous at the
vertices, and that the probability current is conserved, i.e.\ 
\begin{equation}
  \label{eq:Neu2}
  \sum_{v \in b} \frac{\rmd}{\rmd x} \efun_b(v) = 0 \qquad
  \mbox{ for all } v\in\cV,
\end{equation}
where the sum is over all bonds that originate from the vertex $v$ and
the derivatives are taken at the vertex $v$ in the outward direction.
The admissible matching conditions were classified in, among other
sources, \cite{KosSch99, Har00, kuc:qgI}.

A solution to the eigenvalue equation (\ref{eq:Neu2}) on the bond $b$,
can be written as a linear combination of plane waves,
\begin{equation}
  \label{eq:plane-waves}
  \psi_b(x_b) = c_b \rme^{\rmi k x_b} + \hat{c}_b \rme^{-\rmi k x_b} \ .
\end{equation}
A solution on the whole graph can be uniquely defined by specifying
the corresponding vector of coefficients
$\vec{c}=(c_1,\dots,c_B,\hat{c}_1,\dots,\hat{c}_B)^T$.  The elements of the
vector $\vec{c}$ are naturally associated with {\em directed\/} bonds
of the graph $G$.

Imposing the matching conditions, we find that $\spec^2$ is an
eigenvalue if and only if $\S(\spec)\vec{c} = \vec{c}$ for some
explicitly given matrix $\S(\spec)$.  For instance, in the case of
Kirchhoff conditions given above, 
\begin{equation}
  \label{eq:factorzn}
  \S(\spec) = \rme^{\rmi\spec\L}\S_0,  
\end{equation}
where the elements of the $\dim\times \dim$ matrix $\S_0$ are given by
\begin{equation*}
  (\S_0)_{(v_1,v_2)(v_3,v_4)} = \delta_{v_2,v_3} \left(
    \frac{2}{d_{v_2}} - \delta_{v_1,v_4} \right),
\end{equation*}
for $(v_1,v_2)$ and $(v_3,v_4)$ directed bonds of the graph.
By $\L$ we denoted the vector of bond lengths
$\L=(L_1,\ldots,L_\numb)$ and defined, in a slight abuse of notation, a
$\dim\times\dim$ diagonal matrix
\begin{equation}
  \label{eq:abuse}
  \rme^{\rmi \vec{x}}\coloneq\left(
    \begin{array}{cccccc}
      \rme^{\rmi x_1} \\
      & \ddots \\
      & & \rme^{\rmi x_\numb} \\
      & & & \rme^{\rmi x_1} \\
      & & & & \ddots \\
      & & & & & \rme^{\rmi x_\numb}
    \end{array}\right),
\end{equation}
where $\vec{x}\in\R^\numb$.  Note the doubling of dimension which
signifies the move from undirected to directed bonds.

The general conditions for factorization (\ref{eq:factorzn}) are given
in \cite{KPS07,ful:itf}. 
Even when these conditions are not satisfied and $\S$ depends on
$\spec$ in a non-trivial fashion, the above factorization is valid in
the limit $\spec\to\infty$.  Thus, for the purpose of studying the
spectral statistics it is not a strong restriction to assume that 
(\ref{eq:factorzn}) applies, as we will do below.

\subsection{The scattering approach}
The second construction considers wave propagation on the graph where each
vertex is treated as a scatterer and propagation along the bonds is free.
This construction was first considered in the context of studying spectral
statistics in \cite{KS99} and generalized in \cite{Tan00} to directed graphs.
On each bond the waves travel in both directions. Therefore, at any given
moment, the system is fully specified by a $\dim$-dimensional vector of wave
amplitudes $\vec{a}$, indexed again by the directed bonds.  Scattering at
vertices is described by a unitary matrix $\bfT$, having the property that
\begin{equation*}
  T_{(v_1,v_2)(v_3,v_4)} = 0 \qquad \mbox{if} \quad v_2\neq v_3.
\end{equation*}
The free propagation results in the amplitude $a_b$ (here $b$ is a
directed bond) being multiplied by the phase factor $\rme^{\rmi\spec L_b}$.
Altogether we arrive to the {\em quantum evolution operator\/}
$\U(\spec) = \rme^{\rmi\spec\L}\bfT$.

Thus, in both constructions one ends up with a unitary matrix
$\S(\spec) = \rme^{\rmi\spec\L}\S_0$.  This matrix specifies the
eigenvalues $\{\spec_n\}$ {\it via}\/ the equation
\begin{equation}
  \label{eq:sec_eq}
  \det [\Id - \rme^{\rmi\spec\L}\S_0 ] = 0.
\end{equation}

Actually, it was noted in \cite{kur:isp} (see also \cite{GS06,ful:itf}) that
the multiplicity of $\spec=0$ as a root of \eqref{eq:sec_eq} can be different
to its multiplicity as an eigenvalue of the Laplace operator, but that for all
positive eigenvalues, the multiplicities coincide.  To avoid this ambiguity we
will omit the zero eigenvalues from the spectra that we consider. i.e.\ in our
counting, $\spec_1$ will be the first strictly positive eigenvalue.  Since the
multiplicity of the zero eigenvalue is necessarily finite (see corollary
\ref{cor:due} below) this omission will not affect any spectral statistics.

\subsection{Spectral quantities for quantum graphs}
\label{sec:spec_quant}
In various sources the notion of the ``spectrum $\sigma(G)$ of the
graph $G$'' can refer to:-
\begin{enumerate}
\item the eigenproblem (\ref{eq:shrod_eq}) and thus solutions $\{ \spec_n
  \}$ of (\ref{eq:sec_eq}),
\item the eigenphases of the matrix $\S(\spec)$ for an arbitrary
  $\spec$, i.e.  to the set of $\dim$ numbers $\{\theta_j\}$ such that
  $\rme^{\rmi\theta_j}$ is the $j$-th eigenvalue of $\S(\spec)$.
\end{enumerate}
To distinguish the two notions of the spectrum we will refer to the 
first definition above as the $\spec$-spectrum,
and to the second as the $\theta$-spectrum.

In a similar way, the ``eigenvector'' of $G$ can refer to one of three
objects:-
\begin{enumerate}
\item the function $\efun(x)$ that solves (\ref{eq:shrod_eq}), subject to
  boundary conditions, for some $\spec_n$ in the $\spec$-spectrum,
\item the eigenvector of $\S(\spec_n)$ corresponding to the eigenvalue
  1, i.e.\ to the solution $\vec{c}$ of $\S(\spec_n)\vec{c} = \vec{c}$.
  These eigenvectors will be denoted by $\evS_n$,
\item any of the $\dim$ eigenvectors of $\S(\spec)$ for arbitrary $\spec$, 
denoted by
  $\evU_j(\spec)$, $j=1,\ldots,\dim$.
\end{enumerate}
There correspondence between first two types of eigenvectors is given
by formula (\ref{eq:plane-waves}).  A heuristic formula which connects
properties of the second and the third types of eigenvectors was
suggested in \cite[Eq.~(5)]{ber:qeg}.  In this article we give a rigorous
derivation of the formula.

\section{Main results}
\label{sec:results}

Our results concern equivalences between statistical properties of two
types of eigenvalues and eigenvectors of quantum graphs.  These
results are derived as an application of Proposition~\ref{prop:main}
introduced in Section~\ref{sec:tools}.

\subsection{Eigenvalue statistics}
\label{sec:evals}

Starting with the seminal work of Kottos and Smilansky
\cite{KS97,KS99}, quantum graphs became a popular model of quantum
chaos.  One of the more pertinent questions of quantum chaos is the
universality of the eigenvalue correlations among systems of certain
type.  On graphs, \cite{KS97} showed some preliminary numerical
evidence that eigenvalue spectrum of graphs follow the general
prediction \cite{BGS84,cas:otc} which
says that the spectrum of classically chaotic system should have
correlations typical associated to eigenvalues of large random matrices.
Persistent deviations from the predicted behavior were  found in
star graphs \cite{KS99,BK99} and Tanner \cite{Tan01} proposed a
precise condition on the graphs to follow the random matrix theory
prediction.  This question was then attacked analytically by
various methods, with results reported, in particular, in
\cite{BSW02,GA05,ber:ffe}.  For more information we refer the reader to a
recent review \cite{GS06}.

While the original interest was in level statistics, i.e.\ the statistical
functions of the spectrum $\{\spec_n\}$, most of the analytical studies were in
fact concentrating on the eigenphase statistics of the $\theta$-spectrum.
There are heuristic
reasons why the corresponding statistics should coincide, but the formal link
between the two has hitherto not been explored.

In this article we prove that, in the correct limit, the two statistics are
equivalent.  We take the example of nearest-neighbor spacing
distribution, i.e.\ the distribution of gaps $\{\spec_n-\spec_{n-1}\}$ on
one hand and the values of the functions
$\{\theta_j(\spec)-\theta_{j-1}(\spec)\}$ on the other.  Formally, the
level spacing distribution is given by
\begin{equation*}
  P(s) = \lim_{N\to\infty} \frac1{N} \sum_{n=1}^N 
  \delta\big(s - (\spec_n-\spec_{n-1})\big),
\end{equation*}
where $\delta$ is the Dirac delta-function.
Mathematically, the
distributions are defined {\itshape via}\/ a family of test functions.

\begin{theorem}
  \label{thm:spacings}
  Let $\L$ be linearly independent over $\Q$ and $\bar{L} =
  (L_1+\ldots+L_\numb)/\numb$ denote the mean bond length.  If $h$ be a
  continuous function, then the limits
  \begin{equation}
    \label{eq:distro_level}
     P_\spec[h] \coloneq \lim_{N\to\infty} 
     \frac1N \sum_{n=1}^N h\left(\bar{L}(\spec_n-\spec_{n-1})\right),
  \end{equation}
  and
  \begin{equation}
    \label{eq:distro_phase}
    P_\theta[h] 
    = \lim_{\Spec\to\infty} \frac1\Spec \int_0^\Spec  \frac1\dim
    \sum_{j=1}^\dim h\big(\theta_j(\spec)-\theta_{j-1}(\spec)\big)\,\rmd\spec
  \end{equation}
  exist and define bounded linear functionals of $h$.  If we take the limit
  $\L \to \ell_0 (1, 1,\ldots, 1)$ for some $\ell_0>0$, while keeping $\L$
  linearly independent over $\Q$, the two functionals coincide.  Namely,
  \begin{equation}
    \label{eq:limit_equality}
    \lim_{\Delta L\to0} P_\spec[h] = \lim_{\Delta L\to0} P_\theta[h],
  \end{equation}
  where $\Delta L \coloneq
 L_{\mathrm{max}} - L_{\mathrm{min}}$ is the spread of the
  distribution of the bond lengths.
\end{theorem}

\begin{remark}
  If $\L$ is \emph{not} linearly independent over $\Q$, the
  limits~(\ref{eq:distro_level}) and (\ref{eq:distro_phase}) still
  exist but are highly sensitive to the changes in individual bond
  length.  Thus the limits in equation~(\ref{eq:limit_equality}) are,
  in general, singular.  In particular when all bond lengths are equal
  ($\Delta L = 0$), the $\spec$-spectrum is periodic and spacings in
  $\theta$-spectrum are independent of $\spec$.  The nearest-neighbor
  distribution in this case is highly degenerate.
\end{remark}

\begin{remark}
  It is straightforward to extend the result to $r$-th nearest neighbor
  spacing
  distributions, i.e.\ the distributions of $\spec_n-\spec_{n-r}$ and
  $\theta_j(\spec)-\theta_{j-r}(\spec)$.  Moreover, when all $r$-th nearest
  neighbor
  spacing distributions coincide, so do other statistical functions such as
  the $n$-point correlation functions. (See, for example, equation (6.1.39)
  of \cite{meh:rm}.)
\end{remark}

\subsection{Eigenvector statistics}
\label{sec:evecs}

The equivalence between the statistics of the $\spec$-spectrum and the
$\theta$-spectrum can be extended to eigenfunction statistics.

To proceed, we need to introduce some notation in addition to that introduced
previously in section~\ref{sec:spec_quant}.  By $\evU_j(\spec)$ we will denote
the $j$-th eigenvector of $\S(\spec)$.  As before, $\L$ denotes the vector of
the bond lengths.  The total length of the graph is $\curlyL =
\sum_{b=1}^\numb L_b$ and the average bond length, $\curlyL/\numb$, is denoted
by $\bar{L}$.

Let $\A$ be a $\dim\times\dim$ matrix (``observable'').  We denote
$A_n = \langle \evS_n | \A | \evS_n \rangle$ and $A_j(\spec) = \langle
\evU_j(\spec) | \A | \evU_j(\spec) \rangle$, which corresponds to the
``expected value of the observable $\A$''.  We define the bond-length
observable by $L\coloneq \diag(\vec{L},\vec{L})$ and set
$L(\spec_n) = \langle \evS_n | L | \evS_n \rangle$.

Further, let $\D$ denote the $\dim\times\dim$ random unitary matrix
$$D\coloneq \rme^{\rmi \vec{X}},$$
where $\vec{X}$ is a random vector distributed uniformly on
$\Torus\coloneq[0,2\pi)^B$, and $A_j(\D) = \langle
\evU_j(D) | \A | \evU_j(D) \rangle$ be the expectation of $\A$ with
respect to the $j$-th eigenvalue of the matrix $\D\S_0$.  
By $\Daverage{\cdot}$ we denote averages with respect to the 
distribution of $\D$.

One of the central results in the study of statistical properties of
eigenvectors is ``quantum ergodicity''.  Quantum ergodicity is the
property of almost all eigenvectors to equidistribute \cite{sch:epe, zel:ude, 
cdv:eef}.  In our
context, this corresponds to vanishing of the variance of either
$A_n$ or $A_j(\spec)$ in some limit.  We shall not discuss here
the important question of which limit is appropriate; we refer the
interested reader to \cite{ber:qeg}.  Instead we prove that the
variances (as indeed all other moments) of $A_n$ and
$A_j(\spec)$ are equivalent.

\begin{theorem}
  \label{thm:evec_moments}
  Let $\L$ be linearly independent over $\Q$.  Then, for all $m\geq0$,
  \begin{equation}
    \label{eq:var_rel}
    \lim_{N\to\infty} \frac1N \sum_{n=1}^N
    \frac{A_n^m} {L(\spec_n) / \bar{L}}
    = \lim_{\Spec\to\infty} \frac1\Spec \int_0^\Spec 
    \frac1{\dim} \sum_{k=1}^{\dim} A_j(\spec)^m\, \rmd\spec
    = \Daverage{ \frac1{\dim} \sum_{k=1}^{\dim} 
      A_j(\D)^m }.
  \end{equation}
\end{theorem}

The weighting on the left-hand side of \eqref{eq:var_rel} shows that there is
an exact equivalence of moments only in the (additional) limit $\Delta L\to
0$, as was taken in \cite{BK99,ber:nqe}. However, provided $\Delta L$ is
bounded away from zero and infinity, the centered moments would approach the
zero limit either simultaneously or not at all.  Another important feature of
equation~(\ref{eq:var_rel}) is that its second part relates the moments of the
eigenvectors of a fixed graph to the averaged properties of an {\em
  ensemble\/} of random matrices.  While such averaging was not necessary in
the proof of quantum ergodicity of models considered in \cite{ber:qeg}, it is
expected to be helpful in more general circumstances.

\section{Main tools}
\label{sec:tools}

\subsection{Graph spectra and torus flow}
\label{sec:flow}

Our methods are based on an idea going back to Barra and Gaspard
\cite{BarGas00}, who suggested to view the eigenvalues $\spec_n$ as
the Poincar\'e return times of a flow to a hypersurface defined by an
extension of equation~(\ref{eq:sec_eq}).

Let $\Torus$ be the torus of side $2\pi$ in $\numb$ dimensions. 
Define the surface $\Sigma\subseteq \Torus$ by
\begin{equation}
  \label{eq:sigmaset}
  \Sigma \coloneq \left\{ \vec{x} : 
    \det[\rme^{\rmi \vec{x}} \S_0 - \Id] = 0
  \right\},
\end{equation}
where we use notation~(\ref{eq:abuse}) for exponential of a vector.

Further, define a flow $\flow_t(\vec{x}_0)$ on $\Torus$ by
\begin{equation}
  \label{eq:flow_def}
  \flow_t(\vec{x}_0) = \vec{x}_0 + t \L \mod 2\pi.
\end{equation}
If the components of $\vec{L}$ are linearly independent over $\Q$ then
the flow is equidistributing on the torus.

\begin{figure}
\setlength{\unitlength}{5cm}
\begin{center}
\begin{picture}(1,0.9)
\put(0,0){\includegraphics[angle=0,width=7.20cm,height=5cm]{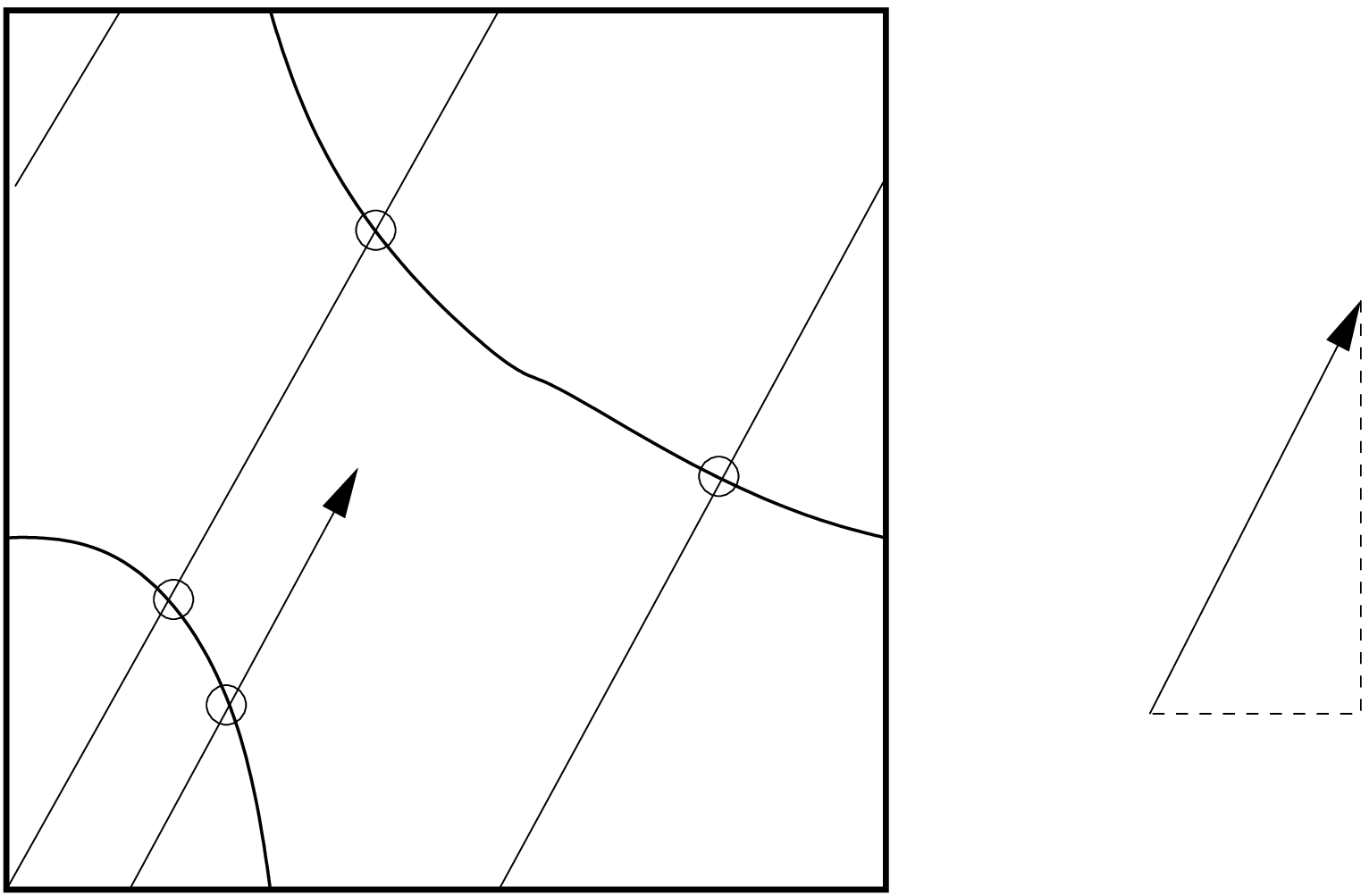}}
\put(0.07,0.30){$\spec_1$}
\put(0.27,0.19){$\spec_4$}
\put(0.29,0.72){$\spec_2$}
\put(0.76,0.36){$\spec_3$}
\put(1.28,0.125){$L_1$}
\put(0.62,0.6){$\Sigma$}
\put(1.44,0.37){$L_2$}
\end{picture}
\end{center}
  \caption{A cartoon of the flow piercing the surface $\Sigma$. The direction
of increasing $k$ is indicated by an arrow, and eigenvalues are indicated 
by little circles.}
  \label{fig:uno}
\end{figure}

Since $\flow_\spec(\vec{0}) = \spec \L \mod 2\pi$,
equation~(\ref{eq:sec_eq}) can be rewritten as $$\det[\rme^{\rmi
  \flow_\spec(\vec{0})} \S_0 - \Id]=0,$$ or simply as
$\flow_\spec(\vec{0}) \in \Sigma$.  Thus the times $t=\spec_n$,
$n\in\Z$ of intersections of the flow $\flow_t(\vec0)$ with the surface
$\Sigma$ give the set of points in the quantum spectrum of a graph
(see figure \ref{fig:uno}).

The idea of Barra and Gaspard \cite{BarGas00} was to use the
ergodicity of the flow (\ref{eq:flow_def}) to compute averages like
(\ref{eq:distro_level}) as an integral over the surface $\Sigma$.  In
this paper we put their idea on a mathematical footing (in particular,
tackling the problem of using the ergodic theorem to integrate over
$\Sigma$ which is measure zero subset of $\Torus$) and extend it to
apply to the problems described in Section~\ref{sec:results}.

\subsection{More about intersections with $\Sigma$}

The following result goes back to at least \cite[Eq.~(70)]{KS99}.  For
completeness, we provide it with a proof.

\begin{lemma} 
  \label{lem:main}
  Let $U$ be a $\dim\times\dim$ unitary matrix, and let $D(t)\coloneq
  \rme^{\rmi \flow_t(\vec{x}_0)}$ for fixed $\vec{x}_0$. Let
  $\rme^{\rmi\theta(t)}$ be an eigenvalue of the unitary matrix $D(t)U$ and
  $\vec{u}(t)$ be the corresponding normalised eigenvector.  Then,
  \begin{equation}
    \label{eq:theta_deriv}
    \frac{\rmd\theta}{\rmd t} = \langle \vec{u}(t) | L | \vec{u}(t) \rangle,
  \end{equation}
  where $L=\diag(\vec{L},\vec{L})$.  In particular, 
  \begin{equation}
    \label{eq:eval:deriv}
    L_{\rm min}\leq \frac{\rmd \theta}{\rmd t}\leq L_{\rm max}.
  \end{equation}
\end{lemma}
\begin{proof}
  Since $D(t)U$ is unitary and analytic, the function
  $\theta(t)$ can be chosen real analytic, and so there are no problems
  differentiating  the eigenvalue equation,
  \begin{equation} \label{eq:deriv:uno}
    D(t)U\vec{u}(t) = \rme^{\rmi \theta(t)}\vec{u}(t),
  \end{equation}
  with respect to $t$ to obtain
  \begin{equation}
    \rmi L D(t)U \vec{u}(t) + D(t)U \vec{u}'(t) = \rmi\theta'(t)\rme^{\rmi
      \theta(t)}\vec{u}(t) + \rme^{\rmi \theta(t)}\vec{u}'(t).
  \end{equation}
  Multiplying by
  $\vec{u}(t)^\dag$ (the conjugate transpose of $\vec{u}(t)$),
  \begin{equation}
    \rmi\rme^{\rmi \theta(t)}\vec{u}(t)^\dag L \vec{u}(t) + \vec{u}(t)^\dag
    D(t)U\vec{u}'(t) = \rmi\theta'(t)\rme^{\rmi \theta(t)},
  \end{equation}
  where we used that $\vec{u}(t)^\dag\vec{u}(t)=1$ and
  $\vec{u}(t)^\dag\vec{u}'(t)=0$.  To handle the second term on the left-hand
  side we notice that
  \begin{equation}
    \vec{u}(t)^\dag D(t)U = \left( (D(t)U)^\dag \vec{u}(t)\right)^\dag 
    = \left( \rme^{-\rmi\theta(t)}\vec{u}(t)\right)^\dag
    = \rme^{\rmi\theta(t)}\vec{u}(t)^\dag
  \end{equation}
  and so 
  $$
  \vec{u}(t)^{\dag} D(t) U \vec{u}'(t) = \rme^{\rmi\theta(t)}\vec{u}(t)^\dag
  \vec{u}'(t) = 0.
  $$
  Thus,
  \begin{equation}
    \frac{\rmd\theta}{\rmd t} = \vec{u}(t)^\dag L \vec{u}(t)
    = \sum_{j=1}^{\numb} L_j (|u(t)_j|^2+|u(t)_{j+\numb}|^2),
  \end{equation}
  leading to the desired bounds.
\end{proof}

We can fix bounds on the rate of crossing of the flow with $\Sigma$ based
on the fact that
\begin{displaymath}
  \det[\rme^{\rmi \flow_t(\vec{x}_0)}\S_0-\Id] = 0
\end{displaymath}
if and only if one of the eigenphases $\theta(t)$ is equal to a multiple of
$2\pi$.  Since all $\theta(t)$ are increasing functions and
Lemma~\ref{lem:main} gives us an estimate of the increase, we get the
following corollary.

\begin{corollary} 
  \label{cor:due}
  If $t$ varies in an interval of length $\displaystyle \frac{2\pi}{L_{\rm
      min}}$ then the flow $\flow_t(\vec{x}_0)$ intersects the surface
  $\Sigma$ \emph{at least} $\dim$ times.

  If $t$ varies in an interval of length $\displaystyle \frac{2\pi}{L_{\rm
      max}}$ then the flow $\flow_t(\vec{x}_0)$ intersects the surface
  $\Sigma$ \emph{at most} $\dim$ times.
\end{corollary}

On average, the flow $\flow_t(\vec{x}_0)$ intersects the surface $2B$ times in
an interval of length $\displaystyle \frac{2\pi}{\bar{L}}$, where $\bar{L}$ is
the {\em average\/} bond length.  This is a consequence of the asymptotic
density of the $\spec$-spectrum, the Weyl's law for graphs.  This classical
result has been known on graphs for some time.  A particularly
elegant proof has appeared recently in \cite[Theorem 1]{gri:muf}.
\begin{proposition}  \label{prop:weyl:law}
  Let $N(\Lambda)\coloneq\#\{ 0< \spec_n \leq \Lambda\}$ where
spectral points are counted with multiplicity. Then
\begin{equation} \label{eq:weyl:law}
  N(\Lambda)\sim \frac{\curlyL}{\pi}\Lambda\qquad
\mbox{as $\Lambda\to\infty$,}
\end{equation}
where $\curlyL\coloneq L_1+\cdots +L_B$.
\end{proposition}
We remark that, in fact, the stronger asymptotic
$$N(\Lambda)= \frac{\curlyL}{\pi}\Lambda + \Ord(1).$$
is proved in \cite{gri:muf}, however we shall only make use of \eqref{eq:weyl:law}.
\subsection{The main proposition}

\begin{figure}
\setlength{\unitlength}{5cm}
\begin{center}
\begin{picture}(1,0.9)
\put(0,0){\includegraphics[angle=0,width=5.0cm,height=5cm]{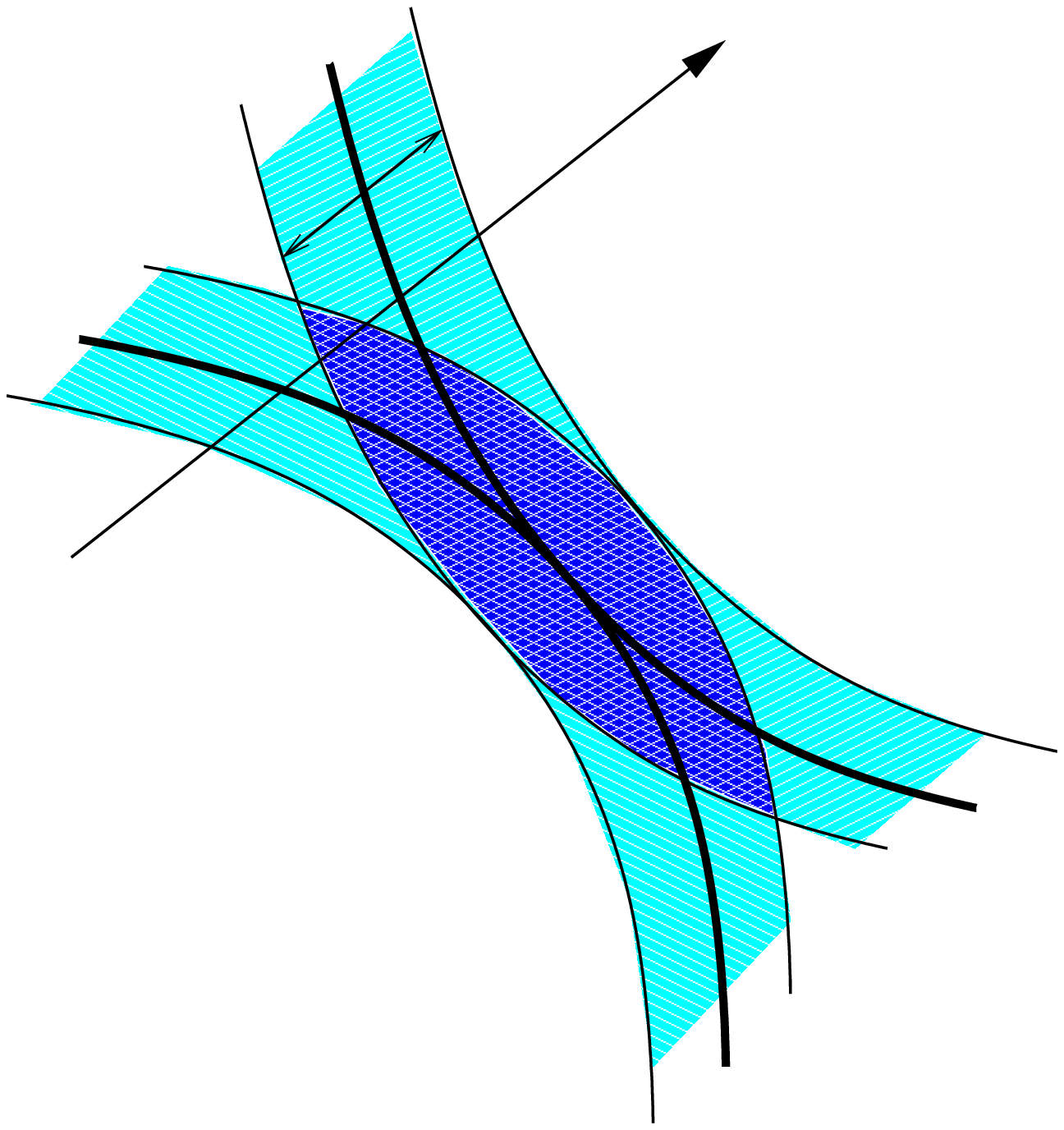}}
\put(0.44,0.89){$\epsilon$}
\put(0.67,-0.03){$\Sigma$}
\put(0.93,0.25){$\Sigma$}
\put(0.73,0.95){Direction of flow}
\end{picture}
\end{center}
  \caption{The region in which the ``thickened'' function $\Phi_\epsilon$ is
    supported. If two branches of $\Sigma$ come close together, the function
    takes the sums of the values on the different branches. In the figure this
    region is shaded in a darker tone.}
  \label{fig:due}
\end{figure}

Let $\Phi$ be a bounded function, defined on $\Sigma$.  We extend it
to the whole of $\Torus$, by the following ``thickening''. Let
$\epsilon>0$ and define
\begin{equation}
  \label{eq:phiepsilon}
  \Phi_\epsilon(\vec{x}) 
  \coloneq \sum_{\vecxi\in\Sigma} \Phi(\vecxi)
  \I_{\{\vecxi=\vec{x}+t\vec{L} \colon |t|\leq\epsilon/2\}}.
\end{equation}
Thus $\Phi_\epsilon$ takes the value $\Phi(\vecxi)$ at all points
within a distance at most $\epsilon/2$ from $\vecxi\in\Sigma$ in the
direction of the flow. Should two or more different branches of the
surface $\Sigma$ come close together, or intersect, the function
$\Phi_\epsilon$ takes the sum of the values of $\Phi(\vecxi)$ on the
branches (see figure \ref{fig:due}).  The sum in \eqref{eq:phiepsilon}
is finite as a consequence of corollary \ref{cor:due}.

We permit the function $\Phi$ to be undefined on a set ${\mathcal U}\subseteq
\Sigma$, provided that ${\mathcal U}$ is small in the following sense.  The
projection of the set ${\mathcal U}$ onto the $\numb-1$ dimensional hyperplane
$P\coloneq\{\x\in\Torus\colon x_1=0\}$
is required to have measure zero:
\begin{equation}
  \label{eq:badset}
  \mathrm{meas} \left\{ \vec{y}\in P : \exists t,\ 0<t\leq2\pi/L_1,\ 
    \flow_t(\vec{y}) \in {\mathcal U}
  \right\} = 0.
\end{equation}
This ensures that the discontinuities of $\Phi_\epsilon$ have measure
zero with respect to $\Sigma$.

Our main tool is the following proposition.

\begin{proposition} \label{prop:main}
  Let $\vec{L}$ be linearly independent over $\Q$ and let $\Phi$ be
  defined on all of $\Sigma$ except possibly for a set ${\mathcal U}$ satisfying
  \eqref{eq:badset}, and be positive and bounded.  Denote by
  $t_n=t_n(\vec{x}_0)$ the consecutive intersection times of the flow
  $\flow_t(\vec{x}_0)$ with the surface $\Sigma$.  Then for any
  $\vec{x}_0\in\Torus$ and $\epsilon>0$
  \begin{equation}
    \label{eq:mainproposition}
    \lim_{N\to\infty}\frac1{N}\sum_{n=1}^N \Phi(\flow_{t_n}(\vec{x}_0))
    = \frac1{\bar{d}(2\pi)^\numb\epsilon}\int_{\Torus} 
    \Phi_\epsilon(\vec{x})\,\rmd\vec{x},
  \end{equation}
  where $\bar{d} \coloneq \lim_{n\to\infty} n/t_n = \curlyL/\pi$ is the
  mean density of the intersections $t_n$.
\end{proposition}

\begin{remark}
  Since we want the result for \emph{every} point
  $\vec{x}_0\in\Torus$, simple ergodicity of the flow $\flow_t$ would
  not be enough.  Instead we shall use the Weyl's equidistribution
  which applies to intervals (and Riemann integrable functions) rather
  than Lebesgue measurable sets (and integrable function).  Namely we use the
  following lemma, which follows directly from the results of Weyl
  \cite{Wey16}.
\end{remark}

\begin{lemma}
  \label{lem:Weyl_equi}
  Let $G\colon \Torus \to \R$ be a Riemann integrable function.  Then,
  for every $\vec{x}_0\in\Torus$,
  \begin{equation}
    \label{eq:weyl_equi}
    \lim_{T\to\infty} \frac1T\int_0^T G(\phi_t(\vec{x}_0))\,\rmd t 
    = \frac1{|{\mathbb T}^\numb|}\int_{\Torus} G(\x) \,\rmd \x.
  \end{equation}
  If the integral on the left-hand side does not exist,
  equation~(\ref{eq:weyl_equi}) should be understood in terms of upper and
  lower Darboux integrals (limits of Darboux sums),
  \begin{equation}
    \label{eq:weyl_equi_Darboux}
    \lim_{T\to\infty} \frac1T 
    \,\mathrm{L.D.}\!\!\!\int_0^T G(\phi_t(\vec{x}_0))\,\rmd t 
    = \lim_{T\to\infty} \frac1T 
    \,\mathrm{U.D.}\!\!\!\int_0^T G(\phi_t(\vec{x}_0))\,\rmd t 
    = \frac1{|\Torus|} \int_{\Torus} G(\x)\, \rmd \x.
  \end{equation}
\end{lemma}

\begin{proof}
  The proof follows the standard procedure, see for example
  \cite[Theorem 1.1]{KuipersNiederreiter}.  Let
  \begin{equation} \label{eq:bigcup}
    \Torus = \bigcup_j R_j
  \end{equation}
  be a finite partition of $\Torus$ into well-behaved sets, such as
  rectangles.  Then
  \begin{displaymath}
    G(\vec{x}) \leq \hat{G}(\vec{x}) \coloneq \sum_j \I_{R_j}(\vec{x}) 
\sup_{\vec{y}\in R_j} G(\vec{y}),
  \end{displaymath}
  where $\I_A$ is the indicator function of set $A$.  By \cite[Satz 5]{Wey16}, 
  \begin{displaymath}
    \lim_{T\to\infty} \frac1T \int_0^T \I_{R_j}(\phi_t(\vec{x}_0))\,\rmd t 
    = \frac{|R_j|}{|\Torus|},
  \end{displaymath}
  the volume of the set $R_j$.  Therefore,
  \begin{displaymath}
    \limsup_{T\to\infty} \frac1T 
    \,\mathrm{U.D.}\!\!\!\int_0^T G(\phi_t(\vec{x}_0))\,\rmd t
    \leq \lim_{T\to\infty} \frac1T \int_0^T \hat{G}(\phi_t(\vec{x}_0))\,\rmd t
    = \frac1{|\Torus|} \sum_j |R_j| \max_{\vec{y}\in R_j} G(\vec{y}),
  \end{displaymath}
  which is the upper Darboux sum for the integral $|{\mathbb T}|^{-\numb}
  \int_{\Torus} G(\x) \rmd \x$.  Since $G$ is Riemann integrable, the Darboux
  sums converge to the value of the integral in the limit as the partition
  \eqref{eq:bigcup} becomes finer.  Analogous lower estimates lead to
  (\ref{eq:weyl_equi_Darboux}).
\end{proof}

\begin{proof}[Proof of Proposition~\ref{prop:main}]
  Since the function $\Phi_\epsilon$ is bounded and the discontinuity set has
  measure zero, $\Phi_\epsilon$ is Riemann integrable and we can apply
  Lemma~\ref{lem:Weyl_equi}.  According to the way that $\Phi_\epsilon$ is a
  thickening of width $\epsilon$ of $\Phi$ on $\Sigma$, we have that
  \begin{equation}
    \label{eq:proof:uno}
    \frac1{N}\sum_{n=1}^N\Phi(\phi_{t_n}(\vec{x}_0)) 
    = \frac1{N\epsilon} \int_0^{t_N+\epsilon/2} 
    \Phi_{\epsilon}(\phi_t(\vec{x}_0))\,\rmd t 
    + \Ord\!\left(\frac1N\right),
  \end{equation}
  as $N\to\infty$.  The error term comes from a possible overlap of the
  integration range with crossings other than $t_n$, $1\leq n\leq N$.
  Corollary~\ref{cor:due} guarantees that the number of such crossings is
  uniformly bounded.
  
  Now we let $N\to\infty$ which entails $T\coloneq t_N+\epsilon/2\to\infty$
  and apply Lemma~\ref{lem:Weyl_equi},
  \begin{align}
    \label{eq:curious}
    \lim_{N\to\infty} \frac1{N}\sum_{n=1}^N\Phi(\phi_{\spec_n}(\vec{x}_0))
    &= \lim_{N\to\infty} \frac{t_N}{N\epsilon} 
   \lim_{T\to\infty} \frac1T \int_0^T \Phi_{\epsilon}(\phi_t(\vec{x}_0))\,\rmd t
    \\ \nonumber
    &= \frac{1}{\bar{d}(2\pi)^\numb\epsilon}
    \int_{\Torus} \Phi_\epsilon(\vec{x})\,\rmd\vec{x}. 
  \end{align}
  Curiously, equation (\ref{eq:curious}) provides a proof that the limit of
  $n/t_n$ exists and is independent of $\x_0$.  Indeed, set $\Phi(\vec{x})
\equiv1$ and
  observe that two other limits in equation~(\ref{eq:curious}) obviously
  exist.  Thus, the limit of the sequence $(n/t_n)$ 
  is the same as for the case $\x_0=0$, 
  which is covered by proposition \ref{prop:weyl:law} 
  giving the limit $\curlyL/\pi$.
\end{proof}

\section{Applications} \label{sec:cinque}

With the tools developed in Section~\ref{sec:tools} we can prove
Theorems~\ref{thm:spacings} and \ref{thm:evec_moments}.  While the proofs
follow the same set of basic ideas, the eigenvector statistics proof is
slightly simpler and thus we present it first.

\subsection{Eigenvector statistics}
\label{sec:appl:vector_stat}

In this section we will prove Theorem~\ref{thm:evec_moments}.  We will
do it by introducing a family of spectra, $\{\spec_{\alpha,n}\}$ indexed
by $\alpha\in[0,2\pi)$.  We remind the reader that $\spec_n$ were defined
as solutions of the equation $\det[\rme^{\rmi \spec\vec{L}}\S_0 - \Id]
= 0$.  We extend this definition and denote by $\spec_{\alpha,n}$ the
solutions of the equation
\begin{equation}
  \label{eq:det_alpha}
  \det[\rme^{-\rmi\alpha} \rme^{\rmi \spec\vec{L}}\S_0 - \Id] = 0.
\end{equation}
Thus, $\spec_{\alpha,n}\vec{L}-\vecalpha \in \Sigma$, where
$\vecalpha=\alpha(1,\ldots,1)$ and $\Sigma$ is the surface defined by
(\ref{eq:sigmaset}).  In other words, $\spec_{\alpha,n}$ are the intersection
times of the flow $\flow_t(-\vecalpha)$ with $\Sigma$.

Similarly to $\evS_n$ we can now define $\evS_{\alpha,n}$ to be the
eigenvector of eigenvalue 1 of the matrix $\rme^{\rmi\alpha}
\rme^{\rmi \spec_{\alpha,n}\vec{L}}\S_0$.  Obviously, $\spec_n =
\spec_{0,n}$ and $\evS_n = \evS_{0,n}$.  We let $A_{\alpha,n} \coloneq \langle
\evS_{\alpha,n} | \A | \evS_{\alpha,n} \rangle$.

\begin{lemma}
  \label{lem:alpha_moment}
  Let $G$ be a bounded continuous and possibly nonlinear functional on
  $\mathbb{C}^{\dim}$.  For $\vec{x}\in\Sigma$ define
  $\Phi(\vec{x})\coloneq G\left(\evS_{\vec{x}}\right)$, where
  $\evS_{\vec{x}}$ satisfies
  \begin{equation*}
    \rme^{\rmi \vec{x}} \S_0 \evS_{\vec{x}} = \evS_{\vec{x}}.
  \end{equation*}
  Then
  \begin{equation} 
    \label{eq:QE:main}
    \lim_{N\to\infty} \frac1N \sum_{n=1}^N G\left(\evS_{\alpha,n}\right) 
    = \frac1{(2\pi)^\numb\epsilon}
    \int_{\Torus} \Phi_\epsilon(\vec{x})\,\rmd\vec{x}.
  \end{equation}
\end{lemma}

\begin{proof}
  First we remark that $\Phi(\vec{x})$ is well-defined and continuous
  on $\Sigma$, except possibly at degenerate points of $\Sigma$.  
  
  By definition, $\evS_{\alpha,n} = \evS_{\vec{x}}$ with $\vec{x} =
  \spec_{\alpha,n} \L - \vecalpha$.  Therefore, 
  \begin{equation*}
    G\left(\evS_{\alpha,n}\right)
    = \Phi\left(\spec_{\alpha,n} \L - \vecalpha\right) 
    = \Phi\left(\flow_{\spec_{\alpha,n}}(-\vecalpha)\right).    
  \end{equation*}
  Now we can apply Proposition~\ref{prop:main} to conclude the proof.
\end{proof}

The important consequence of Lemma~\ref{lem:alpha_moment} is that the
moments of $A_{\alpha,n}$ are independent of $\alpha$.

\begin{proof}[Proof of Theorem~\ref{thm:evec_moments}]

  Let $\left\{\rme^{\rmi \theta_j(\spec)}\right\}_{j=1}^{\dim}$ be the
  eigenvalues of the matrix $\rme^{\rmi \spec \L}\S_0$, with
  $\theta_j(\spec)$ chosen to be real analytic.  As a consequence of
  lemma \ref{lem:main}, $\theta_j$ has a smooth inverse. 
  
  Using a $2\pi$-periodized Dirac delta function, $\delta_{2\pi}$ we can
  trivially write
  \begin{equation*}
    A_j(\spec)^m 
    = \int_0^{2\pi} \delta_{2\pi}\!\left(\alpha-\theta_j(\spec)\right)
    A_j(\spec)^m \,\rmd\alpha.
  \end{equation*}
  Then
  \begin{align}
    \nonumber \int_0^\Spec \sum_{j=1}^{\dim} A_j(\spec)^m\,\rmd\spec 
    &=
    \int_0^{2\pi} \sum_{j=1}^{\dim} \int_0^\Spec
    \delta_{2\pi}\!(\alpha-\theta_j(\lambda))
    A_j(\spec)^m \,\rmd\spec\rmd\alpha\\
    &= \int_0^{2\pi} \sum_{j=1}^{\dim} \int_{\theta_j(0)}^{\theta_j (\Spec)}
    \delta_{2\pi}\!(\alpha-\xi)
    \frac{A_j(\theta^{-1}_j(\xi))^m}{\theta_j'(\theta^{-1}_j(\xi))}
    \,\rmd\xi\rmd\alpha.  \nonumber \\ \label{eq:step1} 
    &= \int_0^{2\pi}
    \sum_{0<\spec_{\alpha,n}\leq\Lambda}
    \frac{A_{j_{\alpha,n}}(\spec_{\alpha,n})^m}
    {\theta_{j_{\alpha,n}}'(\spec_{\alpha,n})} \rmd\alpha,
  \end{align}
  We have chosen $j_{\alpha,n}$ so that
  $\theta_{j_{\alpha,n}}(\spec_{\alpha,n}) = \alpha \mod 2\pi$.  Therefore
  $\evU_{j_{\alpha,n}}\left(\lambda_{\alpha,n}\right)$ is what we
  denoted by $\evS_{\alpha,n}$.
  Further, by Lemma~\ref{lem:main}, $\theta_j'(\spec_{\alpha,n}) = \langle
  \evS_{\alpha,n} | L | \evS_{\alpha,n} \rangle$.  We set 
  \begin{equation*}
    G(\evS) = \frac{\langle\evS|\A|\evS\rangle^m}
    {\langle\evS|\L|\evS\rangle}, 
    \qquad \mbox{so that} \qquad
    \frac{A_{j_{\alpha,n}}(\spec_{\alpha,n})^m}
   {\theta_{j_{\alpha,n}}'(\spec_{\alpha,n})} =
    G\left(\evS_{\alpha,n}\right),
  \end{equation*}
  and obtain
  \begin{align} \nonumber
    \int_0^\Spec \sum_{j=1}^{\dim} A_j(\spec)^m \,\rmd\spec
    &= \int_0^{2\pi} 
   \sum_{\spec_{\alpha,n}\leq\Spec} G\left(\evS_{\alpha,n}\right)\,\rmd\alpha \\
    &= N(\Spec) \int_0^{2\pi}
    \frac1{N(\Spec)} \sum_{\lambda_{\alpha,n}\leq\Spec} 
    G\left(\evS_{\alpha,n}\right) \,\rmd\alpha.
    \label{eq:step2}
  \end{align}
  Now we divide by $\dim\Spec$ and take the limit $\Spec\to\infty$.
  By proposition \ref{prop:weyl:law}, $N(\Spec)/\Spec \to \curlyL/\pi$ in
this limit. By the dominated convergence theorem we can take the limit
  inside the $\alpha$-integral to get
  \begin{align}
    \lim_{\Spec\to\infty} \frac1\Spec \int_0^\Spec 
    \frac1{\dim} \sum_{j=1}^{\dim} A_j(\spec)^m \,\rmd\spec
    &= \frac{\curlyL}{\dim\pi} \int_0^{2\pi} 
    \lim_{\Spec\to\infty} \frac1{N(\Spec)} \sum_{\spec_{\alpha,n}\leq\Spec} 
    G\left(\evS_{\alpha,n}\right) \,\rmd\alpha \nonumber \\
    &= \frac{\curlyL}{B} \lim_{\Spec\to\infty} \frac1{N(\Spec)}
    \sum_{n=1}^{N(\Spec)} G\left(\evS_{0,n}\right) \nonumber \\
    &=  \lim_{N\to\infty} \frac1{N}
    \sum_{n=1}^{N} \frac{A_{n}^m}{L(\spec_{n})/\bar{L}},
    \label{eq:step3}
  \end{align}
  where we used the fact that, by Lemma~\ref{lem:alpha_moment}, the
  limit inside the integral is independent of $\alpha$.  

  To prove the second equality in \eqref{eq:var_rel} we apply Weyl's
  equidistribution theorem to the limit on the left in \eqref{eq:step3}.
  This is justified since eigenvector components of $\rme^{\rmi \vec{x}} S_0$
  vary continuously in $\vec{x}$ except possibly at the points where
  the eigenvalues of $\rme^{\rmi \vec{x}} S_0$ are non-simple.  The set of
  such points is a $C$-analytic set and will have either measure 0 or full
  measure (see the discussion in Section~\ref{sec:generic}).  The latter
  situation is incompatible with condition~(\ref{eq:badset}).
\end{proof}

\subsection{Spectral statistics}
\label{sec:appl:spec_stat}
We will use the previous results to prove that the spectral statistics
of the matrix $\rme^{\rmi \vec{x}}S_0$ averaged over $\Torus$ converge
to the empirical spectral statistics as the interval from which bond
lengths are drawn shrinks in size.

\subsubsection{A few more definitions}
\label{sec:more_defs}

Let $d:\Torus\to\R_+$ be the ``time'' of the next crossing of the
surface $\Sigma$,
\begin{equation}
  d(\vec{x}) \coloneq \inf\{ t>0: \flow_t(\vec{x})\in\Sigma\}.
\end{equation}

For $\vec{x}\in\Torus$ let the eigenvalues of $\rme^{\rmi \vec{x}}S_0$
be $\rme^{\rmi\that_1(\vec{x})},\ldots,\rme^{\rmi\that_{2v}(\vec{x})}$,
ordered so that
\begin{equation}
  \label{eq:theta:ordering}
  0 < \that_{\dim}(\vec{x}) \leq \that_{\dim-1}(\vec{x})
  \leq \cdots \leq \that_2(\vec{x})\leq \that_1(\vec{x}) \leq 2\pi.
\end{equation}
This ordering ensures that if $\vec{x}\in\Sigma$ then $\that_1(\vec{x})=2\pi$.
We use the hat in notation $\that_j$ to distinguish it from the eigenphases
$\theta_j(\spec)$ of the matrix $\rme^{\rmi\spec \L}\S_0$ which can be chosen
real analytic with respect to $\spec$.  Of course, the sets
$\left\{\theta_j(\spec)\right\}$ and
$\left\{\that_j\left(\flow_\spec(\vec{0})\right)\right\}$ coincide modulo
$2\pi$.

\begin{figure}
\setlength{\unitlength}{7cm}
\begin{center}
\begin{picture}(2,1)
\put(0,0){\includegraphics[angle=0,width=14.0cm,height=7cm]{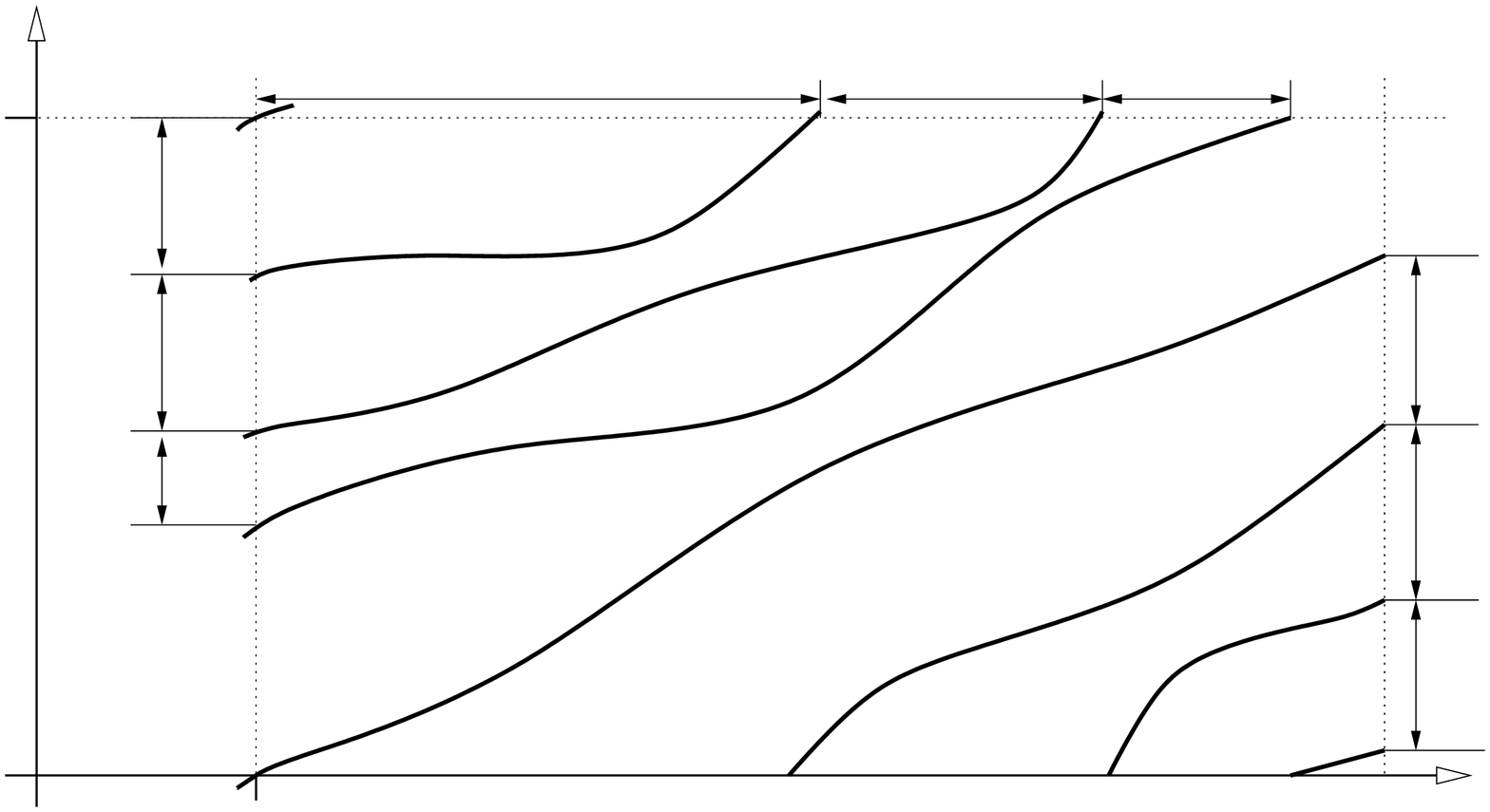}}
\put(0.64,0.92){$d(\x)$}
\put(1.2,0.92){$d(\flow_{\spec_1}(\x))$}
\put(1.52,0.92){$d(\flow_{\spec_2}(\x))$}
\put(-0.07,0.85){$2\pi$}
\put(-0.03,0.03){$0$}
\put(0.07,0.75){$\sigma_1(\x)$}
\put(0.07,0.55){$\sigma_2(\x)$}
\put(0.07,0.40){$\sigma_3(\x)$}
\put(0.3,-0.04){$\spec=0$}
\put(1.06,-0.04){$\spec_1$}
\put(2,-0.03){$\spec$}
\put(1.93,0.57){$\sigma_1(\phi_\lambda(\x))$}
\put(1.93,0.35){$\sigma_2(\phi_\lambda(\x))$}
\put(1.93,0.15){$\sigma_3(\phi_\lambda(\x))$}
\end{picture}
\end{center}
  \caption{The eigenangles of $\rme^{\rmi\flow_\spec(\vec{x})}\S_0$ and
    notation of Section~\ref{sec:more_defs}}
  \label{fig:tre}
\end{figure}

Define the eigenphase spacing functions by
\begin{align*}
  \sigma_1(\vec{x}) &\coloneq \that_1(\vec{x}) - \that_2(\vec{x}) \\
  &\;\;\vdots\\
  \sigma_{\dim-1}(\x) &\coloneq \that_{\dim-1}(\x) - \that_{\dim}(\x) \\
  \sigma_{\dim}(\x) &\coloneq \that_{\dim}(\x) + 2\pi - \that_1(\x).
\end{align*}
These definitions are illustrated in figure \ref{fig:tre}.

We will use shorthand $\sigma_j(\lambda)$ for
$\sigma_j(\flow_\lambda(\vec{0}))$ and $d(\lambda)$ for
$d(\flow_\lambda(\vec{0}))$.

\subsubsection{Equivalence of spacing distributions}

\begin{proof}[Proof of Theorem~\ref{thm:spacings}]
  Let $h$ be a continuous test function.  We want to compare
  \begin{equation*}
    \lim_{\Spec\to\infty} \frac1\Spec \int_0^\Spec  \frac1\dim
    \sum_{j=1}^\dim h\left(\sigma_j(\spec)\right)\,\rmd\spec
    \qquad \mbox{and} \qquad
    \lim_{N\to\infty} \frac1N \sum_{n=1}^N 
    h\left(\bar{L}d(\spec_n)\right).
  \end{equation*}
  
  Repeating the arguments of equations~(\ref{eq:step1}) and
  (\ref{eq:step2}) we obtain
  \begin{align*}
    \int_0^\Spec  \sum_{j=1}^\dim
    h\left(\sigma_j(\spec)\right)\,\rmd\spec
    &= \int_0^\Spec \int_0^{2\pi} \sum_{r=1}^{2B} \delta_{2\pi}\!(\alpha
    - \theta_r(\spec)) h\left(\sigma_{j_r}(\spec)\right)\, 
    \rmd\alpha\rmd\spec \\
    &= \int_0^{2\pi} \int_0^\Spec \delta(\spec-\spec_{\alpha,n}) 
    \frac{h(\sigma_{j_{\alpha,n}}(\spec))}
    {\theta_{j_{\alpha,n}}'(\spec)}\, \rmd\spec\rmd\alpha,\\
    &= \int_0^{2\pi} \sum_{\spec_{\alpha,n}\leq\Spec}
    \Phi^\sigma\left( \flow_{\spec_{\alpha,n}}(-\vecalpha)\right)\,\rmd\alpha,
  \end{align*}
  where, for $\x\in\Sigma$, we have defined
  \begin{equation}
    \label{eq:Phi_sigma_def}
    \Phi^\sigma(\x) = \frac{h\left(\sigma_1(\x)\right)}
    {\left.\frac{\rmd}{\rmd t}\that_1(\flow_t(\x))\right|_{t=0}},
  \end{equation}
  where the derivative is taken from the left, since $\that_1$ is
  discontinuous on $\Sigma$.  The function $\Phi^\sigma(\x)$ is well defined
  if at $\x$ we have $\that_2 < \that_1 = 2\pi$, i.e.\ the eigenvalue 1 of
  $\rme^{\rmi\vec{x}}\S_0$ is simple.  Now we divide by $\dim\Spec$, take the
  limit $\Spec\to\infty$ and use Proposition~\ref{prop:main} to get,
  analogously to (\ref{eq:step3}),
  \begin{align*}
    \lim_{\Spec\to\infty} \frac1\Spec \int_0^\Spec \frac1\dim
    \sum_{j=1}^\dim h\left(\sigma_j(\spec)\right) \,\rmd\spec
    &= \frac{\curlyL}{\dim\pi} \int_0^{2\pi} 
    \lim_{\Spec\to\infty} \frac1{N(\Spec)} \sum_{\spec_{\alpha,n}\leq\Spec}
    \Phi^\sigma\left( \flow_{\spec_{\alpha,n}}(\vecalpha)\right)
    \,\rmd\alpha \\
    &= \frac{\bar{L}}{(2\pi)^\numb\epsilon}
    \int_{\Torus} \Phi^\sigma_\epsilon(\vec{x})\,\rmd\vec{x}.
  \end{align*}
  On the other hand, again by Proposition~\ref{prop:main}, we have
  \begin{equation*}
    \lim_{N\to\infty} \frac1N \sum_{n=1}^N h\left(\bar{L}d(\spec_n)\right)
    = \frac1{(2\pi)^\numb\epsilon}
    \int_{\Torus} \Phi^d_\epsilon(\vec{x})\,\rmd\vec{x},
  \end{equation*}
  where
  \begin{equation}
    \label{eq:Phi_d_def}
    \Phi^d(\vec{x}) = h\left(\bar{L}d(\vec{x})\right).
  \end{equation}
  Now we use that $\bar{L}\Phi^\sigma(\vec{x}) - \Phi^d(\vec{x}) \to 0$ as
  $\Delta L\to0$.  This is proved in lemma~\ref{lem:twospacings} below,
  implying
  \begin{equation}
    \label{eq:Phi_lim_agree}
    \frac{\bar{L}}{(2\pi)^\numb\epsilon}
    \int_{\Torus} \Phi^\sigma_\epsilon(\vec{x})\,\rmd\vec{x} 
    - \frac1{(2\pi)^\numb\epsilon}
    \int_{\Torus} \Phi^d_\epsilon(\vec{x})\,\rmd\vec{x} \to 0.
  \end{equation}
  Since $\Phi^d(\vec{x})$ is clearly continuous in $\L$, the individual limits
  exist too, concluding the proof.  
\end{proof}

\begin{lemma}
  \label{lem:twospacings}
  Let $h$ be continuous.  Then, for $\x\in\Sigma$ we have
  \begin{equation}
    \label{eq:spacings_comp}
    \frac{\sigma_1(\x)}{L_{\rm max}} \leq
    d(\vec{x}) \leq \frac{\sigma_1(\x)}{L_{\rm min}},
  \end{equation}
  and, therefore, 
  \begin{equation}
    \label{eq:Phis_conv}
      \bar{L}\Phi^\sigma(\vec{x}) - \Phi^d(\vec{x}) \to 0
  \end{equation}
  uniformly on $\Sigma$ as $\Delta L = L_{\rm max} - L_{\rm min} \to 0$.
\end{lemma}

\begin{proof}
  Estimate~(\ref{eq:spacings_comp}) follows from lemma
  \ref{lem:main}.  Indeed, $d(\x)$ is the time required by
  $\that_1(\phi_t(\vec{x}))$ to reach $2\pi$ from its initial value of
  $2\pi-\sigma_1+0$ at the time $t=+0$, see Figure~\ref{fig:tre}, and
  lemma~\ref{lem:main} provides an estimate on the derivative of $\that_1$.
  
  We can now write (\ref{eq:spacings_comp}) as
  \begin{equation*}
    \sigma_1(\x)\left(\frac{\bar{L}}{L_{\rm max}}-1\right) 
    \leq \bar{L}d(\vec{x}) - \sigma_1(\x) 
    \leq \sigma_1(\x)\left(\frac{\bar{L}}{L_{\rm min}}-1\right).
  \end{equation*}
  Since both $d$ and $\sigma_1$ are bounded, we have $\Lbar d(\vec{x}) -
  \sigma_1(\x) \to 0$ uniformly in $\x$ as the spread of the lengths
  decreases.  The function $h$ is continuous on a compact set, therefore it is
  uniformly continuous and $h(\sigma_1(\x)) - h(\Lbar d(\vec{x})) \to 0$ 
  uniformly
  in $\x$.  Finally, the denominator in the definition of $\Phi^\sigma$, see
  equation~(\ref{eq:Phi_sigma_def}), is bounded by $L_{\rm min}$ and $L_{\rm
    max}$ by lemma~\ref{lem:main} and thus converges to $\bar{L}$.  All of the
  above put together imply (\ref{eq:Phis_conv}).
\end{proof}

\subsection{Discontinuities of $\Phi^\sigma$ and $\Phi^d$}
\label{sec:generic}

When defining the function $\Phi$ in Section~\ref{sec:appl:vector_stat} and
the functions $\Phi^\sigma$ and $\Phi^d$ in Section~\ref{sec:appl:spec_stat} we
implicitly assumed that they satisfy condition~(\ref{eq:badset}) on the set of
their discontinuities.  In this section we explain why can we expect it in
general and how to rectify the situation when the definitions of function
$\Phi$ produce too many discontinuities.

The surface $\Sigma$ is a $C$-analytic set, see \cite[Definition 6, Chapter
V]{Narasimhan}.  Thus we can apply \cite[Proposition 18, Chapter
V]{Narasimhan}, to it, which splits $\Sigma$ into an analytic
manifold of dimension $\numb-1$ and the remainder ${\mathcal U}$.  
The remainder has a
smaller dimension and obviously satisfies condition~(\ref{eq:badset}).

Consider a connected piece of the $(\numb-1)$-dimensional manifold.  
If on this piece the
zero of the defining equation
\begin{displaymath}
  \det[\rme^{\rmi \flow_t(\vec{x}_0)}\S_0-\Id] = 0
\end{displaymath}
is simple, there is a unique eigenvector $\evS_{\vec{x}}$ (see
Section~\ref{sec:appl:vector_stat}), continuous with respect to $\vec{x}$ on
the manifold.  The function $\Phi$ is well defined.

The situation when on all $(\numb-1)$-dimensional manifolds the eigenvalues are
simple is expected to be generic.  However, a proof exists only for a special
case of graphs with Kirchhoff matching conditions \cite{Fri05}.  Proving the
simplicity (in a generic sense) of the solutions to 
\begin{equation*}
  \det[\rme^{\rmi \x}\S_0-\Id] = 0
\end{equation*}
for general $S_0$ thus remains an important open question.  To complicate the
picture, there are counter-examples with persistent eigenvalues of
multiplicity two: these are provided by graphs with a looping bond separated
from the rest of the graph.  Moreover, there are indications that the
first-order perturbation used in \cite{Fri05} will not produce the result for
general $\S_0$.

Fortunately, when the multiplicity is greater than one on a
$(\numb-1)$-dimensional submanifold of $\Sigma$, we can still make the
functions $\Phi$ continuous.  First we note that the multiplicity must be
constant on the whole manifold.  For simplicity we assume the multiplicity is
two.  The eigenspace of the eigenvalue $1$ of the matrix $\rme^{\rmi
  \flow_t(\vec{x}_0)}\S_0$ is continuous and we can choose a continuous basis
$\evS_{\vec{x},1}$ and $\evS_{\vec{x},2}$.  The function $\Phi$ of
Section~\ref{sec:appl:vector_stat} can be then defined as
\begin{displaymath}
  \Phi(\vec{x}) = G\left(\evS_{\vec{x},1}\right) 
  + G\left(\evS_{\vec{x},2}\right)
\end{displaymath}
on the problematic manifold.

The function $\Phi^d$ of Section~\ref{sec:appl:spec_stat} can be simply
defined as $h(\Lbar d(\x)) + h(0)$.  To define function $\Phi^\sigma$ in a
continuous and meaningful manner we note that if $\that_1(\x) = \that_2(\x) =
2\pi$, then the phases also coincided before the flow hit $\Sigma$, i.e. for
small negative $t$, $\that_1(\flow_t(\x)) = \that_2(\flow_t(\x))$.  Thus the
derivative in the denominator of equation~(\ref{eq:Phi_sigma_def}) does not
depend on which $\that$ we take and one can set
\begin{displaymath}
  \Phi^\sigma(\x) = \frac{h\left(\sigma_1(\x)\right) + h(0)}
  {\left.\frac{\rmd}{\rmd t}\that_1(\flow_t(\x))\right|_{t=0}},
\end{displaymath}

\subsection*{Acknowledgments}
We are grateful for a number of interesting discussions with
J.~Bolte, D.~Grieser, J.~Harrison, J.P.~Keating and P.~Kuchment.

This work is supported by the National Sciences Foundation under
research grant DMS-0604859.

\bibliographystyle{brianbib2}
\bibliography{scatterbib}
\end{document}